\documentclass[twocolumn, showpacs, prl]{revtex4}

\usepackage{graphicx}% Include figure files

\usepackage{dcolumn}% Align table columns on decimal point

\usepackage{bm}% bold math

\usepackage{color}

\begin{document}

\title{Parallel magnetic field tuning of valley splitting in AlAs two-dimensional electrons}

\date{\today}

\author{T.\ Gokmen}

\author{Medini\ Padmanabhan}

\author{O.\ Gunawan}

\author{Y. P.\ Shkolnikov}

\author{K.\ Vakili}

\author{E. P.\ De Poortere}

\author{M.\ Shayegan}

\affiliation{Department of Electrical Engineering, Princeton
University, Princeton, NJ 08544}

\begin{abstract}

We demonstrate that, in a quasi-two-dimensional electron system
confined to an AlAs quantum well and occupying two conduction-band
minima (valleys), a parallel magnetic field can couple to the
electrons' orbital motion and tune the energies of the two valleys
by different amounts. The measured density imbalance between the two
valleys, which is a measure of the valley susceptibility with
respect to parallel magnetic field, is enhanced compared to the
predictions of non-interacting calculations, reflecting the role of
electron-electron interaction.

\end{abstract}


\maketitle

The physics governing the valley splitting in multi-valley
two-dimensional electron systems (2DESs), such as in Si
field-effect-transistors, has been of interest for some time
\cite{Dorda78PRB, Ando82RevModPhys}. It is attracting renewed
attention in 2DESs in Si, AlAs, and graphene
\cite{Eng07PRL,Geim2007Nature}, both as a fundamental problem, and
also because of the possibility that manipulating the electron
valley degree of freedom might lead to novel ("valleytronics")
devices \cite{Gunawan2007PRB,Gunawan2008PRL,Rycerz2007Nature}.
Traditionally, the valley energies have been controlled via strain,
confinement and electric field \cite{Dorda78PRB, Ando82RevModPhys,
Takashina2004PRB, ShkolnikovFootnote}. In this Letter we demonstrate
how a magnetic field ($B_{\parallel}$) applied $parallel$ to the 2D
plane can also break the valley degeneracy and shift the valley
energies of a 2DES with finite layer thickness. In such a system,
$B_{\parallel}$ couples to the electron orbital motion and deforms
the electron wave function in the confinement direction. To first
order, this deformation increases the valley energies (diamagnetic
shift) and, to second order, it increases the effective mass in the
direction perpendicular to $B_{\parallel}$ \cite{Batke87PRB,
Kunze87PRB, Stern67, Stern68, Tutuc03PRB}. In our AlAs samples,
where two valleys with anisotropic Fermi contours are occupied, when
$B_{\parallel}$ is applied along the major axis of one the valleys
and perpendicular to the other valleys' major axis, it shifts the
valleys' energies by different amounts. Remarkably, the measured
energy shift and the resulting valley density imbalance are much
larger than simple calculations would predict, signaling a clear
enhancement of the $B_{\parallel}$-induced valley spitting, likely
due to electron-electron interaction. (The parameter $r_{s}$,
defined as the average inter-electron spacing measured in units of
the effective Bohr radius, ranges from 6.4 to 9.8 for our samples.)

\begin{figure}
\centering
\includegraphics[scale=1.0]{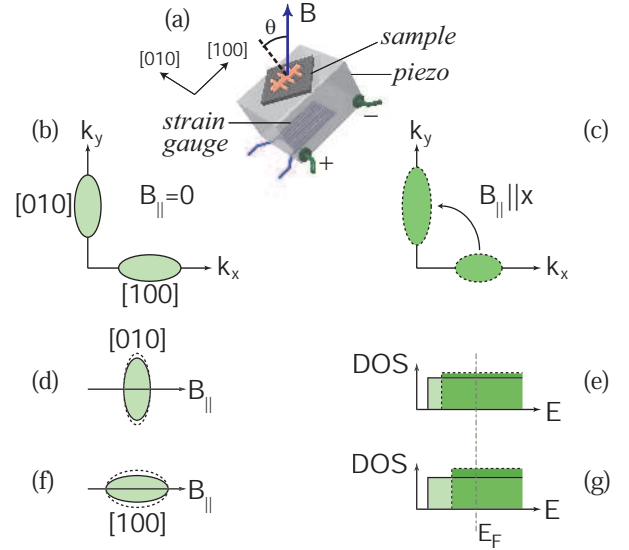}
\caption{(Color online) Schematic diagrams showing: (a) Experimental
setup; (b) and (c) Valley occupations at zero (light green) and
finite (dark green) parallel magnetic fields; (d) and (f) Fermi
contour distortions due to parallel magnetic field for the [100] and
[010] valleys; (e) and (g) Density-of-states (DOS) for the [100] and
[010] valleys before (light green) and after (dark green) the
application of parallel magnetic field.}
\end{figure}

Figure 1 summarizes our experimental setup and measurements. We
studied four samples; here we focus on two samples where the 2DES is
confined to either an 11 nm-wide or a 15 nm-wide AlAs quantum well
(QW), grown using molecular beam epitaxy on a semi-insulating GaAs
(001) substrate. The AlAs well is flanked by AlGaAs barriers and is
modulation-doped with Si \cite{PoortereAPL2002}. In these samples
the electrons occupy two conduction-band valleys with elliptical
Fermi contours as shown in Fig. 1(b) \cite{ShayeganPSS2006}, each
centered at an X point of the Brillouin zone, and with an
anisotropic effective mass (longitudinal mass $m_{l}=1.05$ and
transverse mass $m_{t}=0.20$, in units of free electron mass,
$m_{e}$) \cite{Gunawan2007PRB}. We denote these two valleys
according to the direction of their major axis: [100] and [010]
(Fig. 1(b)). In our experiments $B_{\parallel}$ is applied along
[100] as schematically shown in Fig. 1(a). Since the two valleys
have different orientations with respect to $B_{\parallel}$, the
diamagnetic shift and the effective mass enhancement are different
for the two valleys, as illustrated in Figs. 1(d-g), causing an
electron transfer from the [100] to the [010] valley (see Fig.
1(c)). This density imbalance ($\Delta n$) caused by $B_{\parallel}$
can be countered by applying symmetry-breaking strain $\epsilon
=\epsilon _{[100]} - \epsilon _{[010]}$, where $\epsilon_{[100]}$
and $\epsilon_{[010]}$ are the strain values along the [100] and
[010] directions. In our study, we monitored the sample resistance
vs. $\epsilon$ at different values of $B_{\parallel}$ to determine
$\Delta n$ as a function of $B_{\parallel}$. In order to apply
tunable strain our samples were glued to a piezoelectric stack
actuator as shown in Fig. 1(a)
\cite{ShayeganPSS2006,GunawanPRL2006}. The measurements were
performed in a $^3$He cryostat with a base temperature of 0.3 K. The
system was equipped with a tilting stage, allowing the angle
$\theta$ between the sample normal and the magnetic field to be
varied \textit{in situ}.

Before presenting the experimental data, we will first outline the
theoretical formalism that describes this valley splitting. Using
the same approach as in Refs. \cite{Stern67} and \cite{Stern68}, we
start with a simple, single-particle effective mass Hamiltonian in
three-dimensions:
$$
H_{3D}=\frac{p_x^2}{2m_x}+\frac{(p_y+qB_{\parallel}z)^2}{2m_y}+\frac{p_z^2}{2m_z}+V(z)
\eqno(1)
$$
where $p_x$, $p_y$, $p_z$ are momentum operators, $m_x$, $m_y$,
$m_z$ are effective masses in $x$, $y$, $z$ directions, $q$ is the
electron charge and $V(z)$ is the confinement potential in the $z$
direction which is assumed to be a QW with 210 meV barrier heights
and 11 nm or 15 nm of well width for our two sample structures
\cite{ShayeganPSS2006}. We use the gauge $ \vec{A} = (0,
B_{\parallel}z, 0)$. Solutions to the above Hamiltonian are plane
waves in $x$ and $y$ directions. Substituting the plane wave
solutions leads to the one-dimensional Hamiltonian:
$$
H_{1D}=\frac{p_z^2}{2m_z}+V(z)+\frac{2\hbar k_yqB_{\parallel}z +
(qB_{\parallel}z)^2}{2m_y}\eqno(2)
$$
where the last two terms containing $B_{\parallel}$ can be
described as perturbations to the Hamiltonian
$H_0=p_z^2/2m_z+V(z)$. To first order, the ground state energy is
shifted to higher values because of the second perturbation term,
$(qB_{\parallel}z)^2 / 2m_y$. To second order, the effective mass
in the $y$ direction ($\hbar^2/(d^2E/dk_y^2)$) becomes enhanced
since the first perturbation term ($\hbar k_yqB_{\parallel}z /
m_y$) is linear with $k_y$. Eventually, the total energy for the
ground state can be written as follows:
$$
E=\frac{\hbar^2 k_x^2}{2m_x}+\frac{\hbar^2 k_y^2}{2m_y^{'}}+E_0+
\Delta E_0 \eqno(3)
$$
where $E_0$ is the ground state energy of $H_0$, $\Delta E_0$ is the
ground state energy shift, and $m_y^{'}$ is the enhanced effective
mass in the $y$ direction.

Initially, the two valleys have the same solutions to $H_0$
because of their same mass in the $z$ direction. The crucial
observation is that the perturbation terms in Eq. (2) depend on
$m_y$ which is different for the two valleys. The [100] valley has
a smaller mass in the $y$ direction and therefore feels a stronger
perturbation at a given $B_{\parallel}$. Its energy is shifted
more compared to the [010] valley, causing a splitting with
$B_{\parallel}$. The quantity we measure experimentally is $\Delta
n$ caused by $B_{\parallel}$, and contains two terms: one, the
difference in the ground state energies and the other, the
difference in the density-of-states caused by the mass
enhancement. To second order in perturbation, $\Delta n$ can be
written as:
$$
\Delta n = \alpha (\frac{1}{m_l}-\frac{1}{m_t}) (1-\frac{2 \pi
\hbar^{2}n}{(E_1-E_0) \sqrt{m_l m_t}}) \frac{\sqrt{m_l m_t}}{2 \pi
\hbar^{2}} \eqno(4)
$$
where $\alpha=(1/2)q^2 B_\|^2(\langle z^2 \rangle - \langle z
\rangle ^2)$ and $E_1$ is the first excited state energy of $H_0$.
Although second order perturbation theory gives analytic answers,
in our calculations we numerically solved the Schrodinger's
equation using a finite difference method; i.e., the values we
report are numerically exact solutions to $H_{1D}$ in Eq. (2).

\begin{figure}
\centering
\includegraphics[scale=0.9]{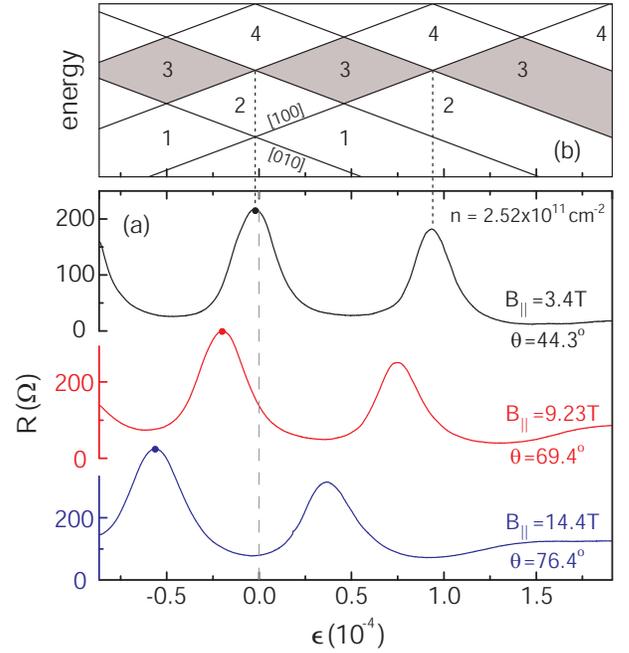}
\caption{(Color online) (a) Resistance vs. strain traces for an 11
nm-wide AlAs QW taken at $\nu=3$ for different parallel magnetic
fields. The "balanced point" for a given $B_{\parallel}$ is marked
by a closed circle on the corresponding trace. (b) Landau level fan
diagram for $B_{\parallel}=3.4$ T as a function of $\epsilon$.}
\end{figure}

Experimentally, we determine $\Delta n$ vs. $B_{\parallel}$ by
monitoring the sample resistance ($R$) as a function of $\epsilon$
in tilted magnetic fields. Examples of such piezo-resistance
traces are shown in Fig. 2(a). Each trace was taken at a fixed
$\theta$ and magnetic field so that the 2DES remains at a fixed
Landau level (LL) filling factor ($\nu$=3 in the case of Fig. 2(a)
data). With applied $\epsilon$, the LLs for the [100] and [010]
valleys cross each other, as the fan diagram in Fig. 2(b)
indicates. $R$ exhibits minima as the Fermi energy ($E_{F}$)
passes through consecutive energy gaps, and maxima as it coincides
with the LL crossings. The fan diagram in Fig. 2(b) is drawn for a
fully spin polarized system; this is indeed the case for the
traces shown in Fig. 2(a) because of the large Zeeman splitting
due to the high magnetic field. From the piezo-resistance traces,
the "balanced point," i.e., the strain at which the valleys are
equally occupied at a given $B_{\parallel}$ can be measured by
following the resistance peaks indicated by closed circles. (The
balanced point at $B_{\parallel}=0$, which corresponds to
$\epsilon=0$, is determined experimentally from $R$ vs. $\epsilon$
sweeps at even fillings as has been detailed in Ref.
\cite{GunawanPRL2006}.) The data of Fig. 2 clearly show that the
valley energies are split with the application of $B_{\parallel}$.

In Fig. 3(a) we show a set of $R$ vs. $\epsilon$ traces taken at a
fixed $\theta$ for $\nu=$ 4, 5 and 6. Again, in all these traces,
the 2DES is fully spin polarized because of the very large
$B_{\parallel}$. The traces are periodic in $\epsilon$, implying
that they are consistent with a simple, linear LL fan diagram, an
example of which is shown in Fig. 6(b) for $\nu=6$. By associating
the $R$ maxima with the LL coincidences, we can determine the
valley polarization, $P_{V}$, defined as the difference between
the [010] and [100] valley populations divided by the total 2DES
density (i.e., $\Delta n/n$), at each coincidence. We therefore
obtain a direct measure of $P_{V}$ vs. $\epsilon$ which we plot,
for $\nu=6$, as black squares in Fig. 3(c); note that $P_{V}$ is
equal to -1/6, 1/6, 3/6 and 5/6 for $\nu=6$ coincidences from left
to right. The lines in Fig. 3(c) provide two pieces of useful
information. First, their intercepts with the $P_{V}$ axis provide
direct measures of $P_{V}$, or equivalently, $\Delta n$, for given
values of $B_{\parallel}$. In the inset to Fig. 3(c), we show a
plot of $\Delta n$ vs. $B_{\parallel}^2$. Note that the plot is
approximately linear, consistent with what we expect from Eq. (4).
Second, the slopes of the lines in Fig. 3(c) give a measure of the
valley susceptibility with respect to strain
\cite{GunawanPRL2006}, i.e., the rate of change of $\Delta n$ with
$\epsilon$; note that the slopes are independent of $\nu$ for a
fixed density, consistent with the measurements of
\cite{GunawanPRL2006} which were done in the absence of parallel
field.

\begin{figure}
\centering
\includegraphics[scale=0.90]{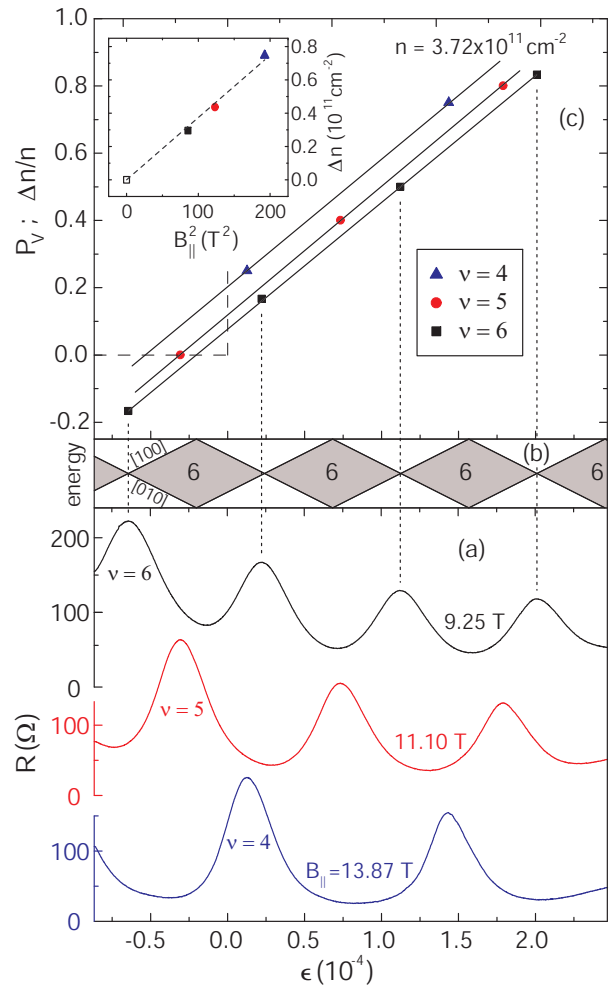}
\caption{(Color online) (a) Resistance vs. strain traces for the
same sample as in Fig. 2 but at a higher density, taken at different
fillings for a fixed tilt angle, $\theta=74.5^{\circ}$. (b) Landau
level fan diagram as a function of $\epsilon$ for $\nu=6$. (c)
Valley polarization ($P_{V}$) as a function of $\epsilon$ at
different fillings. The intercepts of the lines with the $y$ axis
give $P_{V}$ (or alternatively, $\Delta n$) for the corresponding
values of $B_{\parallel}$. Inset shows $\Delta n$ as a function of
$B_{\parallel}^2$.}
\end{figure}

\begin{figure*}
\centering
\includegraphics[scale=1.0]{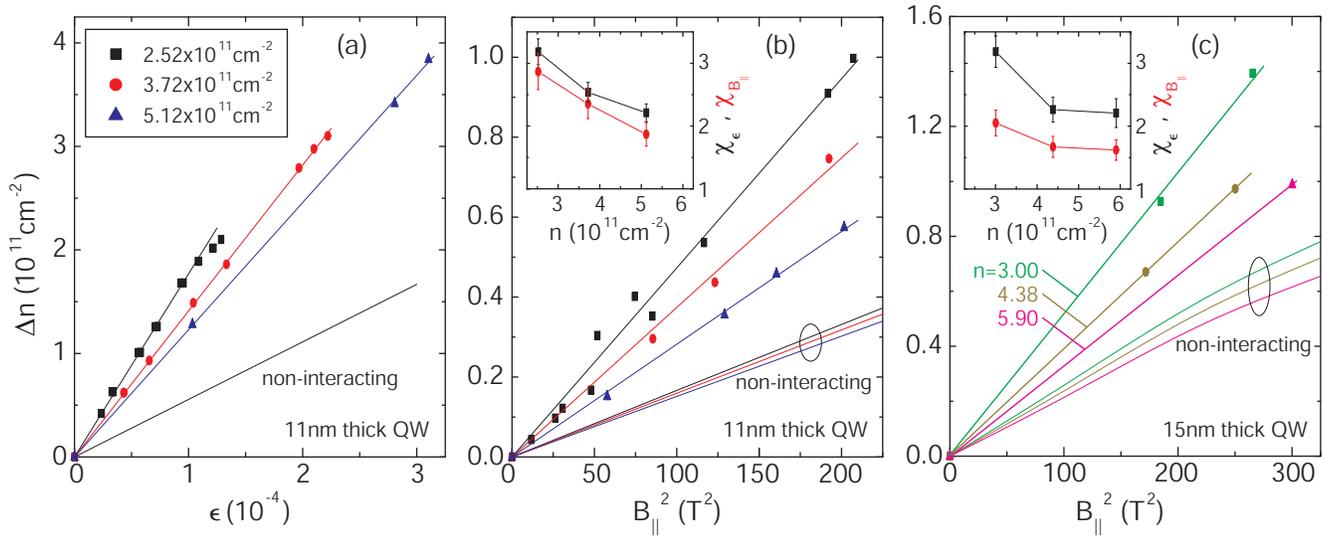}
\caption{(Color online) (a) Summary of density imbalance as function
of $\epsilon$ for three densities for the 11 nm-wide QW. (b) and (c)
Summary of density imbalance as function of $B_{\parallel}^2$ for
three densities for the 11 nm-wide and 15 nm-wide AlAs QWs. Insets
show the valley susceptibilities, normalized to their band values,
with respect to two variables, $\epsilon$ and $B_{\parallel}^2$.}
\end{figure*}

To summarize our results, in Figs. 4(a) and (b) we plot the measured
$\Delta n$ vs. $\epsilon$ and vs. $B_{\parallel}^2$, respectively,
for the 11 nm-wide AlAs QW at three different densities. The data in
Fig. 4(a) were obtained from lines such as those shown in Fig. 3
(c), but after subtracting their intercepts with the $\Delta n$
axis. Figure 4(a) data essentially represent $\Delta n$ vs.
$\epsilon$ in the absence of $B_{\parallel}$; this can be verified
from comparison of the data with measurements reported in Ref.
\cite{GunawanPRL2006} which were done at $\theta=0$. In Fig. 4(a) we
also include a plot of $\Delta n$ vs. $\epsilon$ expected based on
the band parameters, i.e., ignoring electron-electron interaction
and using the simple relation $\Delta n = \epsilon E_{2} m / 2 \pi
\hbar^2$ where $m=\sqrt{m_l m_t}=0.46$ and $E_{2}$=5.8 eV is the
deformation potential for the AlAs X-point conduction band minimum
\cite{ShayeganPSS2006}. In Fig. 4(a), it is clear that the response
of the system to strain is two to three times enhanced compared to
what the non-interacting (band) parameters predict, and that the
enhancement is stronger at lower densities (larger $r_{s}$). This
observation confirms the enhancement of valley susceptibility with
respect to strain ($\chi_{\epsilon}$), defined as the rate of change
of $\Delta n$ with $\epsilon$, originally reported in Ref.
\cite{GunawanPRL2006}; it is similar to the enhancement of the spin
susceptibility observed in similar samples and reflects the role of
electron-electron interaction
\cite{ShkolnikovPRL2004,GunawanPRL2006,GokmenPRB2007}.

Our measured $\Delta n$ caused by $B_{\parallel}$ are plotted in
Fig. 4(b) vs. $B_{\parallel}^2$ for the same three densities as in
Fig. 4(a). Again, for comparison, we also show the results of the
calculations based on the non-interacting picture, i.e., using Eq.
(4). As can be seen in Eq. (4), there is a slight dependence of
$\Delta n$ on $n$ and this is why there are three lines in Fig. 4(b)
representing the non-interacting calculations \cite{footnote2}.
Similar to the strain case of Fig. 4(a), the system's response to
$B_{\parallel}^2$ is about two to three times stronger than the
non-interacting calculations predict. This is a noteworthy result as
it demonstrates that the interacting 2DES responds to two very
different stimuli (strain and $B_{\parallel}$) in a very similar
fashion.  Defining a new quantity, $\chi_{B_{\parallel}}$, as the
valley susceptibility with respect to $B_{\parallel}^2$, we see from
the inset to Fig. 4(b) that $\chi_{\epsilon}$ and
$\chi_{B_{\parallel}}$ have very similar magnitudes
\cite{footnote1}.

In Fig. 4(c) we show the summary of $B_{\parallel}$-induced
$\Delta n$ for a second sample, a 2DES confined to a 15 nm-wide
AlAs QW. The data overall are quite similar to the 11 nm-wide QW
data. The larger well-width of the 15 nm-wide sample predicts a
larger $\Delta n$ for this sample (at a given density) because of
a larger spread of the wave function in the $z$ direction (see Eq.
(2)). The deduced, normalized susceptibilities plotted in Fig.
4(c) inset, however, appear to be slightly smaller than those for
the 11 nm-wide sample. It is likely that this is also related to
the larger electron layer thickness: Although there have been no
systematic and detailed studies of the dependence of
$\chi_{\epsilon}$ enhancement on electron layer thickness, it has
been reported that the $spin$ susceptibility of a thicker electron
system is less enhanced compared to a thinner system with
otherwise the same parameters \cite{DePaloPRL2005,GokmenPRB2007}.

Our results demonstrate that a magnetic field applied parallel to an
AlAs 2DES with finite layer thickness can split the energies of the
two in-plane valleys. The splitting originates from the coupling of
the parallel field to the orbital motion of the electrons. Our
measurements of the density imbalance due to strain and parallel
magnetic field show similar enhancements of the splitting compared
to calculations which are based on non-interacting electrons with
band parameters. These enhancements are more pronounced at lower
densities (larger $r_{s}$). The results suggest that in an
interacting 2DES, the valley splitting is enhanced by interaction,
independent of the physical parameter that induces the splitting.

We thank the NSF for support, and R. Winkler for useful discussions.
Part of this work was done at the NHMFL, Tallahassee, which is also
supported by the NSF. We thank E. Palm, T. Murphy, J. Jaroszynski,
S. Hannahs and G. Jones for assistance.

\break


\begin{references}
\bibitem{footnote2}
As can be seen from Eq. (4), the mass enhancement leads to a
decrease in $\Delta n$. For example, for the 11 nm-wide sample, the
diamagnetic shift and the mass enhancement lead to $\Delta n$ of
$5.79\times10^{10}$cm$^{-2}$ and $-0.88\times10^{10}$cm$^{-2}$,
respectively at $n=5.12\times10^{11}$cm$^{-2}$ and
$B_{\parallel}=18$T. For the 15 nm-wide sample, these values are
$9.10\times10^{10}$cm$^{-2}$ and $-2.56\times10^{10}$cm$^{-2}$, at
$n=5.90\times10^{11}$cm$^{-2}$ and $B_{\parallel}=18$T.
\bibitem{footnote1}
We define $\chi_{B_{\parallel}}$ as the rate of change of $\Delta
n$ with $B_{\parallel}^2$, and both $\chi_{\epsilon}$ and
$\chi_{B_{\parallel}}$ in Figs. 4(b) and 4(c) insets are
normalized to their band values.
\end{references}
\end{document}